\documentclass[mathpazo]{cicp}

\begin{document}
\title{Time-evolution of nanoscale systems by finite difference method}

\author[M.~Nakhaee]{Mohammad Nakhaee$^{1,\dag}$\corrauth , S Ahmad Ketabi$^{1,\ddag}$, M Taher Pakbaz$^{1,\P}$,M Ali M Keshtan$^{2,\star}$, Elham Rahmati$^{1,\S}$ and Zahra Abdous$^{3,\ast}$}

\address{$^{1}$Damghan University, Damghan, Iran
, $^{2}$Department of Physics, Iran University of Science and Technology, Narmak, Tehran 16844, Iran
, $^{3}$Islamic Azad Universiry Tehran Center Branch, Tehran, Iran}

\email{{\tt m.nakhaee@std.du.ac.ir} (M.~Nakhaee)}
 

\begin{abstract}
Using finite difference method, time evolution of a typical metal-molecule-metal system is studied by 
introducing a new method to solve general related Volterra $integro-differential\,equation$ ($IDE$).
Discretization in time domain is applied for one dimensional chain tight binding model in several cases 
by defining a $matrix\,integro-differential\,equation$ ($MIDE$). Results are compatible with their 
analytical counterparts and show more accuracy than other numerical methods like Runge Kutta ($RK$).
Charge transport properties in a trans-polyacetylene chain are found by studying the time evolution 
of charge density in it and current-voltage diagram is calculated.
\end{abstract}


\maketitle

\section{Introduction}
\label{sec1}
Recently, Metal-Molecule-Metal ($MMM$) structures (Figure 1)  have attracted 
scientists. Their vast applications, include electrical, optical, mechanical, $etc$  have led 
scientists to produce devices with new abilities and have improved efficiencies relative to 
their primary counterparts~\cite{Nitzan,Liang,MMM1,MMM2,MMM3}. Well predicting behaviour of $MMM$ systems 
requires investigating time-evolution of their transport properties. Many efforts have been done 
to investigate time dependence of transport properties in $MMM$ structures. Generally, Green's function 
formalism and density functional theory, have extensively applied to study time evolution of 
$MMM$ structures~\cite{Td1,Td2,Td3,Td4,Td5,Td6,Td7}. In the Green's function formalism, transport properties of $MMM$ systems can 
be deduced by applying Green's function in the energy representation, $G(E)$, which can 
be calculated as follows:
\begin{equation}\label{gr}
G(E)=[(E+i\,0^{+})I-H-\Sigma(E)]^{-1}
\end{equation}
In which $H$, $\Sigma(E)$ and $E$ are Hamiltonian of the molecule and self-energy of the system 
in energy $E$, respectively. The Fourier transform of $G(E)$, suggests impulse response 
$(G(t))$ as:
\begin{equation}\label{grt}
G(t)=\frac{1}{2\pi\hbar}\int_{-\infty}^{\infty} e^{\frac{-iEt}{\hbar}}\,G(E)\,dE 
\end{equation}
which satisfies the Fourier transform of equation \eqref{gr} as follows:
\begin{equation}\label{ftgr}
(i\hbar \frac{\partial}{\partial t}-H-\Sigma)\,G(t)= I\,\delta(t)
\end{equation}
Taking the energy dependence into account, the product of $\Sigma$ and $G$ becomes a convolution 
in time domain then equation \eqref{ftgr} can be rewritten as ~\cite{RefDatta}:
\begin{equation}\label{cftgr}
(i\hbar \frac{\partial}{\partial t}-H)\,G(t)-\int \Sigma(t-t')\,G(t') 
\,dt'=I\,\delta(t)
\end{equation}

Equation~\eqref{cftgr} is a non-homogeneous Volterra IDE \cite{Volterra} with an intractable and time consuming general 
solution process.

Finite difference method ($FDM$), is an applicable scheme to solve coupled equations\cite{FD} 
and as will be mentioned in section \ref{sec:level22}, discretization of a differential 
equation results some ones. This method has been used to study electronic transport in 
nanostructures, for instance Khomyakov et. al. \cite{RefKhomyakov} have calculated coherent 
transport of a nano wire by wave functions matching in the boundary zones connecting electrods 
and the scattering region using $FDM$.  

In this article, using $FDM$, a simple formalism is proposed to solve equation~\eqref{cftgr}. 
In this approach, derivative and integrator operators are defined then equation~\eqref{cftgr} 
is rewritten in matrix form in the presence of adequate boundary conditions (section
\ref{sec:level22}).
    \begin{figure}[ht]\label{Fig-molecule}
    \begin{center}
    \includegraphics[scale=1]{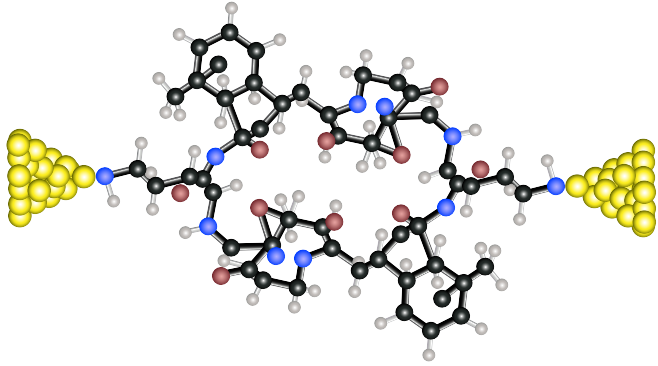}
    \end{center}
    \caption{(Color online) A schematic figure of a metal-molecule-metal system.} 
    \end{figure}    

In section \ref{sec:level4} this approach is applied to calculate  the Green's function, $G(t)$ for an infinite $1D$ chain. Results are calculated in both time-independent and 
time-dependent Hamiltonian cases and are compared with their analytical solutions. Beside 
some numerical comparisons are made between computational errors of our formalism and Dyson 
series and $RK$ methods. Finally, this method is used to calculate the charge current in a 
system composed of a trans-polyacetylene molecule connected to two semi infinite 
$1D$ metal electrodes.

\section{Method}
\label{sec2}
\subsection{\label{sec:level2}Construction of the Finite Difference Scheme to solve the Volterra ${\bf IDE}$}

To solve equation~\eqref{cftgr} generally, consider following Volterra $IDE$:
    \begin{equation}\label{IDE}
    \frac{d}{dt}y(t) = f(t)y(t) + \int_0^t k(t-\tau)y(\tau)d\tau
    \end{equation}
In which $y(t)$ and $f(t)$ stand for functions of an arbitrary real parameter $t$ and 
$k(t-\tau)$ represents a convolution between $\tau$ and $t$. In order to perform numerical 
calculations it is really lucrative to make a discretization scheme for this $IDE$ which allow 
us to use $FDM$. The grid used for discretization is a set of points $\{1,2,\cdots,n_t\}$ where 
$n_t$ is an integer and shows the number of mesh points in the $t$ domain which may be 
determined properly based on the fluctuations of the functions. Commonly, $y_t$, $f_t$ and 
$k_t$ are $n_t$ dimensional vectors which contain all information of functions $y(t)$, $f(t)$ and 
$k(t)$, respectively. Beside we need two operators; a first order derivative operator and an 
integrator one; which are represented by $D^{n_s}_{n_t \times n_t}$ and $I^{n_s}_{n_t \times 
n_t}$, respectively. $n_s$ returns to $n_s$-point stencil of a point in the grid in $FDM$ 
formalism; the point itself together with its $n_s-1$ neighbours. Clearly $n_s$ must usually 
choose in such a way that be less than or equal to $n_t$, ($n_s \le n_t$). Here we try to introduce these two 
operators properly.

The first derivative of a function $y(t)$ respect to the parameter $t$ at a point $t_i$ is 
usually approximated using a $n_s$-point stencil as\cite{pointns}:  
    \begin{equation}\label{difre}
     y'(t_i) \thickapprox d_{i,i} y(t_i) + \sum\nolimits_{i_s \in neighbors} d_{i,i_s} 
     y(t_{i_s})
    \end{equation}

The coefficients $d_{i,j}$  of this equation, while $i$ and $j$ can be integer numbers in this 
set $\{1,2,\cdots,n_t\}$, are well known as Lagrange interpolation coefficients  
\cite{lagrange} and are used widely in $FDM$. These coefficients should be exploited 
to derive $D^{n_s}_{n_t \times n_t}$ as a matrix whose elements are zero except those that 
are $d_{i,j}$. Up to this precision, it is straightforward to define the inverse of this 
derivative operator as an adequate integrator operator, namely $I^{n_s}_{n_t \times n_t}$. 
Uniqueness of this integrator operator imposes a boundary condition. Keeping in mind the 
proper initial value condition, which is $\int_0^0 y(t) dt=0$, we pursue common procedure in 
the $FDM$. For each boundary condition a row and a column are added to integrator matrix 
\cite{FD} so that the final integrator operator with boundary conditions is
introduced as:       
     \begin{equation}\label{integro}
      I^{n_s}_{(n_t+1) \times (n_t+1)}= \left(
      \begin{array}{ccc}
      D^{n_s}_{n_t \times n_t}  & B_{n_t \times 1}  \\
      B^\dagger_{n_t \times1 } &  0 \\
      \end{array}
      \right)^{-1}
     \end{equation}    
In which, the elements of the matrix $B_{n_t \times 1}$ are defined as:     
    \begin{equation}
     B_{t , 1} = \delta_{t,1}
    \end{equation}
In which $\delta_{t,1}$ represents the Kronecker delta function. After inversion, we omit 
added row and column of integrator operator with boundary conditions,
$I^{n_s}_{(n_t+1)\times (n_t+1)}$, and reshape it to new one, namely $I^{n_s}_{n_t\times n_t}$ 
which is integrator operator without boundary conditions. In addition to bringing forward these two operators, we need to shed light on the right hand 
side of the equation \eqref{IDE} and its meaning in our method. For the first part we propose a 
new operator which multiplies $y(t)$ by $f(t)$ and represent it with $M_{n_t \times n_t}$. 
Clearly it must multiply the same elements of $y_t$ by $f_t$. It is satisfied by Exploiting the Kronecker delta function as follows:
    \begin{equation}
     M_{t , t'} = f_{t'} \delta_{t,t'}
    \end{equation}
In the second part, we must integrate a function which is multiplication of the convolution 
function $k(t-\tau)$ by $y(\tau)$. Hiring the predefined integrator operator and representing a 
new integrator operator with convolution factor by $J^{n_s}_{n_t \times n_t}$, we sufficiently 
introduce a matrix whose elements are defined as follow: 
     \begin{equation}
      J^{n_s}_{t,t'}=k_{t-t'+1} I^{n_s}_{t,t'}
     \end{equation}
In which $t$ and $t'$ go from 1 to $n_s$. We can rewrite equation~\eqref{IDE} in matrix form using the operators $ D^{n_s}_{n_t \times n_t} $, $ M_{n_t \times n_t} $ and $J^{n_s}_{n_t \times n_t}$:
    \begin{equation}\label{IDEMF}
    (D^{n_s}-M-J^{n_s}).y=0
    \end{equation}
Where in it subscripts are omitted for abbreviation and matrices are multiplied in usual 
matrix product law. 
\subsection{\label{sec:level3}Construction of the Matrix Finite Difference Scheme to solve  
${\bf MIDE}$}
To extend our method over matrix domain, retaining their definitions, we replace the functions 
$y$, $f$ and $k$ with $Y_{n \times n}$, $F_{n \times n}$ and $K_{n \times n}$, respectively 
where $n$ is the dimension of these matrices. The $IDE$ substitutes by its matrix counterpart 
$MIDE$ as follows:
    \begin{equation}\label{MIDE}
    \frac{d}{dt}Y_{n \times n} (t) = F_{n \times n} (t)Y_{n \times n} (t) + \int_0^t K_{n 
    \times n} (t-\tau)Y_{n \times n} (\tau)d\tau
    \end{equation}
All of the matrices are time dependent. Discretization of these matrices in $t$ domain 
eventuates $3D$ array version of them, namely $Y_{x,x',t}$, $F_{x,x',t}$ and $K_{x,x',t}$ where 
$x$, $x'$ and $t$ are integer numbers. $x$ and $x'$ belong to set $\{1,2,\cdots,n\}$ and $t$ 
goes from 1 to $n_t$ as pointed before. Supporting matrix representation, we reshape 
$Y_{x,x',t}$ to $Y_{\xi,x'}$, in which $\xi$ sweeps both parameters $x$ and $t$. Consequently 
it becomes a $(n_t.n)\times (n)$ matrix. For consistency $\xi$ is defined as:       
      \begin{equation}
       \xi(t,x)  = t + (x-1) n_t
     \end{equation}

By means of our previous definitions, we construct a new first order derivative and an 
integrator operator in this scope. Let $\mathfrak{D}^{n,n_s}_{(n_t.n)\times (n_t.n)}$ denotes 
the first order derivative operator whose elements are defined as: 
     \begin{equation}\label{matr_d}
      \mathfrak{D}^{n,n_s}_{\xi(t,x),\xi'(t',x')}=\delta_{x,x'}D^{n_s}_{t,t'}
     \end{equation}
Where $D^{n_s}_{t,t'}$ is the related matrix entry of $D^{n_s}_{n_t\times n_t}$ which was 
formerly defined in equation \eqref{difre} and $\delta_{x,x'}$ is the Kronecker delta function. 
Suppose $\mathfrak{I}^{n,n_s}_{(n_t.n) \times (n_t.n)}$ stands for the integrator operator. The 
elements of this operator are determined as: 
     \begin{equation} \label{matr_i}
     \mathfrak{I}^{n,n_s}_{\xi(t,x),\xi(t',x')}= K_{x,x',t-t'+1}  I^{n_s}_{t,t'}
     \end{equation}
In this equation $I^{n_s}_{t,t'}$ represents the proper entry of $I^{n_s}_{n_t\times n_t}$,      
once was defined in equation \eqref{integro}. Pursuing our procedure we need an adequate 
operator to multiply $F_{x,x',t'}$ by $Y_{\xi,x'}$. Let $\mathfrak{M}_{(n_t.n)\times(n_t.n)}$ 
represents it. Exerting the Kronecker delta function its elements are assigned as: 
     \begin{equation} \label{SecondPart}
      \mathfrak{M}_{\xi(t,x),\xi(t',x')} = F_{x,x',t'} \delta_{t,t'}
     \end{equation}

We put equations \eqref{matr_d}, \eqref{matr_i} and \eqref{SecondPart} in to the equation 
\eqref{MIDE} to achieve its $FDM$ counterpart as: 
     \begin{equation}
      (\mathfrak{D}^{n,n_s} - \mathfrak{M} - \mathfrak{I}^{n,n_s})Y_{(n_t.n) \times (n)} = 
      \bold{0}_{(n_t.n)\times(n)}
     \end{equation}
In which $\bold{0}_{(n_t.n)\times(n)}$ denotes a zero matrix. Generally $Y(t)$ at $t=0$ is $Y(t=0)=Y_1$, i.e.: 
the $Y_1$ is an arbitrary $n\times n$ matrix at $t=0$ with adequate conditions based on our 
problem. Finally, we obtained first order linear partial integro-differential equation as a 
system of linear equations. To digest we get 
$\mathfrak{A}=\mathfrak{D}^{n,n_s}-\mathfrak{I}^{n,n_s}-\mathfrak{M}$.
     \begin{align}\label{SystemLinear}
      \left(
      \begin{array}{cc}
      \mathfrak{A}_{(n_t.n) \times (n_t.t)}  & {B}_{(n_t.n)\times n} \\
      {B^\dagger}_{(n_t.n)\times n} &  \bold{0}_{n\times n} \\
      \end{array}\right)&\quad
      \left(\begin{array}{c}
      Y_{(n_t.n)\times n} \\
      X_{n \times n} \\
      \end{array}\right)\nonumber\\&= 
      \left(
      \begin{array}{c}
      \bold{0}_{(n_t.n)\times n}\\
      {Y_1}_{n\times n}\\
      \end{array}\right)
    \end{align}
Where $B$ imposes boundary condition at $t=0$ by following definition:
     \begin{equation}
      B_{\xi(x,t),x'}=\delta_{x,x'}\delta_{t,1}
     \end{equation}
In which $\delta_{x,x'}$ and $\delta_{t,1}$ are the Kronecker delta function. It should be 
noted that $X$ is the discarded part of the answer after solving the equation 
\eqref{SystemLinear}. The remainder of this paper is reserved for material to exploit the above 
method.

\section{Numerical results}
\label{sec3}
\subsection{Studying the electronic transport of a one dimensional system }
\begin{figure}[ht]
    \begin{center}
    \includegraphics[scale=0.3]{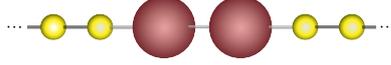}
    \end{center}
    \caption{(Color online) Schematic illustration of a simple $MMM$ consists of two-atom 
    molecule connected to two semi infinite $1D$ leads.} \label{Fig-molecule1}
    \end{figure}

We start with a simple toy model. As a primarily system consider a simple $MMM$ consists of 
two-atom molecule connected to two semi infinite $1D$ leads. As a primarily system, consider an 
infinite one dimensional chain of atoms which some part of it may counts as center part and its 
two tails as two semi infinite $1D$ electrodes(see figure \ref{Fig-molecule1}).
In the tight-binding approximation and second quantization representation, Hamiltonian for 
the molecule can be written as follows:
     \begin{equation}\label{EqHamiltonian}
      H(t)=\sum_x \epsilon_0 c_x^\dagger c_x + \sum_{<x,x'>}^{'} t_{x,x'}(t) c_x^\dagger c_{x'}
     \end{equation}
where the summations run over the lattice sites, $\epsilon_0$ is the energy of the electrons at 
site $x$, $t_{x,x'}$ is the transfer energy between site $x$ and $x'$, and $c^\dagger_x$ 
($c_{x'}$) is the creation (annihilation) operator of electrons at site $x$ ($x'$). The prime 
on summation symbol omits the cases $x=x'$. $\Sigma(t)$ should be determined to solve equation 
\eqref{cftgr}. In this method, $\Sigma(t)$ accounts for the ``interaction'' of an open system; 
namely the center part; with the attached two ideal semi infinite leads. So the non-vanishing 
elements of the self-energy matrix (for this considered system, the first and the last 
elements) will be the conventional self-energy of an ideal semi infinite $1D$ chain which in 
the energy representation can be written as\cite{Sigma}:
\begin{equation} \label{EqSigmaE}
\Sigma(\epsilon)=\left\lbrace\begin{array}{ll}
\frac{v^2}{2}[V-\epsilon-i \sqrt{4-(\epsilon-V)^2}]&|\epsilon-V|<2 \\ \\
\frac{v^2 }{2}[V-\epsilon+\frac{(\epsilon-V)}{|\epsilon-V|}\sqrt{(\epsilon-V)^2-4}]&|\epsilon-
V|>2\end{array}\right.             
\end{equation}
In which $V$ is the bias voltage, $v$ is the coupling energy between leads and the central molecule 
and $\epsilon$ is the energy. $\Sigma(t) $ will be 
obtained by the Fourier transform of the $\Sigma(\epsilon)$:
     \begin{equation} \label{EqSigmat}
      \Sigma(t)=-i v^2\,\Theta(t)\,\frac{J_1(2 t)}{t}\,e^{i V t}
     \end{equation}
where $\Theta(t)$ is the Heaviside function, $J_1$ is the Bessel function of the first kind and other parameters are similar to their
Fourier transforms.\\

Rewriting the $MIDE$ form of the equation \eqref{cftgr} for this system, it will be found:  
    \begin{align}\label{ToyModelEQ}
    i \hbar \frac{d}{dt}&G(t) - \left(
      \begin{array}{cc}
      \epsilon_0  & t_{1,2}(t) \\
      t_{2,1}(t) &  \epsilon_0 \\
      \end{array}
      \right)G(t) \nonumber\\&- \int_0^t \left(
      \begin{array}{cc}
      \Sigma(t-\tau)  & 0  \\
      0  & \Sigma(t-\tau)  \\
      \end{array}
      \right)G(\tau) d\tau = I\,\delta(t)
    \end{align}    
Instead of the Green's function, we continue our approach with time evolution operator $U(t)$ 
which has a simple relation with $G(t)$: [$G(t)=-i \Theta(t) U(t)$]. The boundary condition for 
$U(t)$ at $t=0$ is $U(0)=\hat{1}_{2 \times 2}$. Finally, to use the $FDM$ for 
equation~\eqref{ToyModelEQ}, some substitutions by replacing $Y$ and $Y_1$ in 
equation~\eqref{SystemLinear} with $U$ and $\hat{1}_{2 \times 2}$, respectively and turning $K$ 
and $F$ in equations~\eqref{matr_i} and~\eqref{SecondPart} to self-energy and Hamiltonian 
matrices. For numerical calculations, except for mentioned cases, parameters $n_s=11$ and 
$n_t=1000$ are fixed to obtain time evolution operator (TEO) in the certain domain of $t$; 
($t=[0,\cdots,50]$ in this work)\\

\begin{figure}[ht]
    \begin{center}
    \includegraphics[width=\columnwidth]{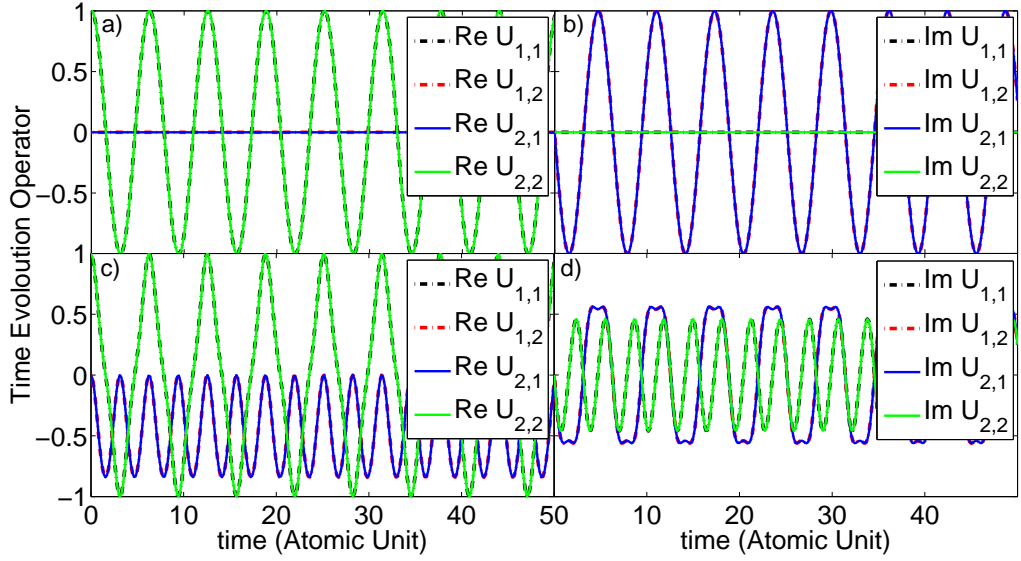}
    \end{center}
    \caption{Real (a and c), and imaginary (b and d), parts of TEO elements for an isolated system, in the cases of time independent (a and b) and a time 
    commutative time dependent (c and d) Hamiltonians, as mentioned in equations \eqref{MatExp} 
    and \eqref{MatExpInt}, respectively.}
      \label{Fig-C12}
    \end{figure}

Two major cases are distinguished, a MMM system in the limit of very weak electrods coupling 
(isolated molecule), which means no convolution term in equation~\eqref{ToyModelEQ} [$v=0$ in 
equations~\eqref{EqSigmaE} and ~\eqref{EqSigmat}] and a traditional MMM system [non-zero 
convolution term in equation~\eqref{ToyModelEQ}]. The simplest model may consist of two atoms 
as the central molecule in which the on-site energy of the electrons ignored and the hopping 
terms are constant [$\epsilon_0=0$ and $t_{x,x'}(t)=1.0$ in the 
equation~\eqref{EqHamiltonian}]. For the isolated system, the exact analytical solution of the 
equation \eqref{ToyModelEQ} for $U(t)$ can straightforwardly be found as:
    \begin{equation} \label{MatExp}
      U(t)=e^{-i H t} = \left(
      \begin{array}{cc}
      cos(t)  & -i sin(t) \\
      -i sin(t) & cos(t) \\
      \end{array}
      \right)
     \end{equation}
In this case the $FDM$ solutions for real and imaginary parts of $U(t)$ are shown in figures 
\ref{Fig-C12} (a) and (b), respectively where in them $U_{i,j}$ stands for the $i^{th}$ and 
$j^{th}$ entry of $U(t)$ matrix. Clearly $FDM$ results for elements of $U(t)$ matrix are in 
excellent agreement with ones which obtained from analytical solutions. 

The presence of a periodic time dependent perpendicular electric field, makes a time 
commutative time dependent Hamiltonian with dynamic on-site energies [$\epsilon_0(t)=cos(t)$]. 
Again, The exact analytical solution of the equation~\eqref{ToyModelEQ} 
for $U(t)$ in this case may be calculated as:
     \begin{equation} \label{MatExpInt}
      U(t)=e^{-i \int_0^t H(\tau) d\tau} = e^{-i sin(t)} \left(
      \begin{array}{cc}
      cos(t)  & -i sin(t) \\
      -i sin(t) & cos(t) \\
      \end{array}
      \right)
     \end{equation}
The $FDM$ solutions for real and imaginary parts of $U(t)$ are shown in 
figures~\ref{Fig-C12} (c) and (d). Once more, analytical results are consistent with $FDM$ 
answers for elements of $U(t)$ matrix superbly.

\begin{figure}[h]
    \begin{center}
    \includegraphics[width=\columnwidth]{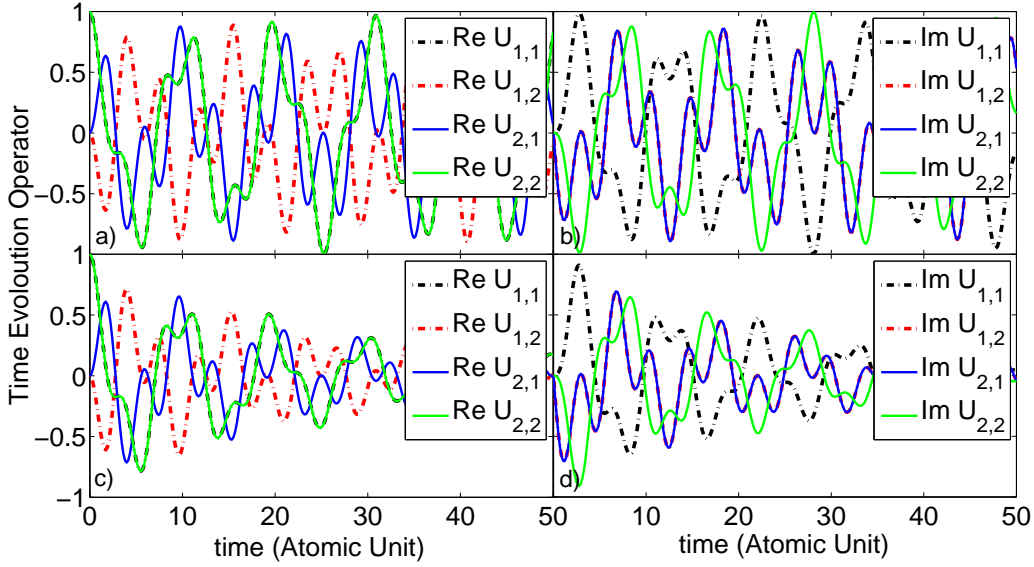}
    \end{center}
    \caption{Real (a and c), and imaginary (b and d), parts of TEO 
    elements for an isolated (a and b) and an interacting (c and d) systems, in the case of 
    time dependent non-commutative Hamiltonian in time, as mentioned in equations 
    \eqref{MatTtExpInt} and \eqref{ToyModelEQ}, respectively.}\label{Fig-C34}
    \end{figure}

In the case of time dependent, non-commutative Hamiltonian in time, a simple case with 
vanishing on-site energies; $\epsilon_0(t)=0$ and time dependent phase hopping 
$t_{x,x'(t)}=e^{-i\Phi(t)}$ are studied. This phase hopping may emerge from an external 
magnetic field or strain, $etc$~\cite{Peierls}. Regardless of the physical source of this 
phase, to facilitate the calculation, a simple time dependent function $\Phi(t)=\Omega t$ with 
$\Omega=1.0$ is selected. Dyson series are usual method to study this types of 
Hamiltonian~\cite{Schiff} which in compact form, exploiting time order operator $T_{\tau}$ to 
respect the time order, it can be noted as~\cite{Fetter}:
      \begin{equation} \label{MatTtExpInt}
      U(t)=T_{\tau}(e^{-i \int_0^t H(\tau) d\tau}) 
     \end{equation}
Although this equation is not as tractable as its former counterparts, it can be solved 
analytically for this specific Hamiltonian and so its answer will be found as:  
       \begin{align} \label{ExactC3}
      U(t)&= \left(
      \begin{array}{cc}
      A(t)&C(t)\\\\
      -C^*(t)&A^*(t)\end{array}\right)\\
      A(t)&=\frac{1}{10}[ (5-\sqrt{5})e^{-i\frac{\sqrt{5}+1}{2}t}+ 
      (5+\sqrt{5})e^{i\frac{\sqrt{5}-1}{2}t} ]\nonumber\\
      C(t)&=-\frac{2i}{\sqrt{5}}\,e^{\frac{it}{2}}\sin(\frac{\sqrt{5}t}{2})\nonumber
     \end{align}
Using $FDM$, numerical solution of the equation~\eqref{MatTtExpInt} for this 
specific Hamiltonian eventuated to figures~\ref{Fig-C34} (a) and (b) for real and imaginary 
parts of the elements of TEO $U(t)$.

In all three types of systems the elements of the TEO are not only periodic in time but also 
non-dissipative which the later one is a natural property of an isolated system. This may be 
counted as a good evidence for authenticity of the procedure.      

For an ``interacting'' system which is more interesting, equation~\eqref{cftgr} contains 
the convolution term. A simple model may be constructed with vanishing on-site energies; 
$\epsilon_0(t)=0 $ and time dependent hopping $t_{x,x'(t)}=e^{-i\Phi(t)}$ with $\Phi(t)=\Omega 
t$ for central molecule and a weak coupling probability between molecule and electrodes; 
$v=0.2$ (as described in third former isolated system). Unfortunately analytical solution for 
this system is intractable but its numerical solution using $FDM$ led to calculate real and 
imaginary parts of $U(t)$ (figures~\ref{Fig-C34} (c) and (d), respectively).

Dissipative behaviour of elements of the TEO may be regarded as a good physical proof for 
``interacting'' nature of the system.

\begin{figure}[h]
    \begin{center}
    \includegraphics[scale=0.3]{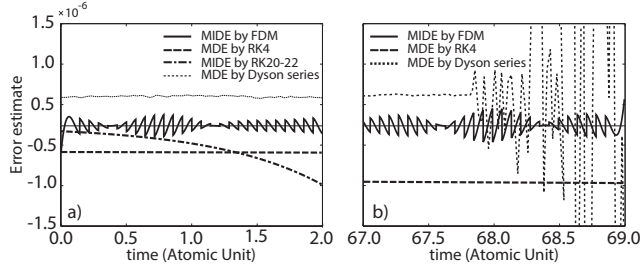}
    \end{center}
    \caption{Error estimate of different methods related to a short period a) and a long enough   
    period b) of time.} \label{Fig-Error}
    \end{figure}

To illustrate both advantages and deficiencies of the $FDM$, comparison of it with 
Dyson series method and $RK$ family methods is appropriate. Figure \ref{Fig-Error} (a) shows 
the error estimates for four distinct conditions for a short period of time after its initial 
condition $t_0=0$. In an isolated system ``without any interactions with electrode''; i.e. 
[$v=0$ in equations~\eqref{EqSigmaE} and ~\eqref{EqSigmat}]; the $IDE$ equation~\eqref{cftgr} 
is modified to a trivial partial differential equation($PDE$) so $RK$ family methods may be the 
best choice to solve it since $RK$ family methods are faster and more accurate than other ones 
[dashed line in figure \ref{Fig-Error} (a)]. Turning the ``interactions'' on, a general 
Volterra $IDE$ [equation~\eqref{cftgr}] must be solved. Because the $RK$ method is a very rough 
approximation to calculate an integral, this method deviates rapidly [dashed dot line in
figure~\ref{Fig-Error} (a)]. Power series characteristic of Dyson series method in time causes 
it to fluctuate in time faster than $FDM$. Comparison of dot line which stands for Dyson series 
method with solid line which represents the $FDM$ diagrams in figure~\ref{Fig-Error} (a) can 
clarify it. Therefore in the vicinity of initial condition in time, both Dyson series method 
and $FDM$ are applicable but the former one is more accurate. After a long enough period of 
time, effect of the higher power of time in Dyson series causes larger fluctuations and more 
deviations in its solution and only $FDM$ can be tractable and accurate. All of them are 
illustrated in figure \ref{Fig-Error} (b).  
\subsection{${\bf FDM}$ application to study electronic transport in a trans-polyacetylene 
molecule }
\begin{figure}[h]
    \begin{center}
    \includegraphics[scale=0.3]{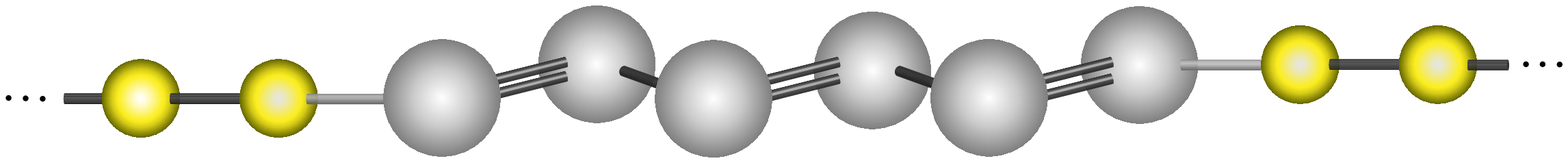}
    \end{center}
    \caption{(Color online) A sketch of the polyacetylene molecule between two semi infinite 
    $1D$ electrodes.} \label{Fig-molecule2}
    \end{figure}

\begin{figure}[h]
    \begin{center}
    \includegraphics[scale=0.4]{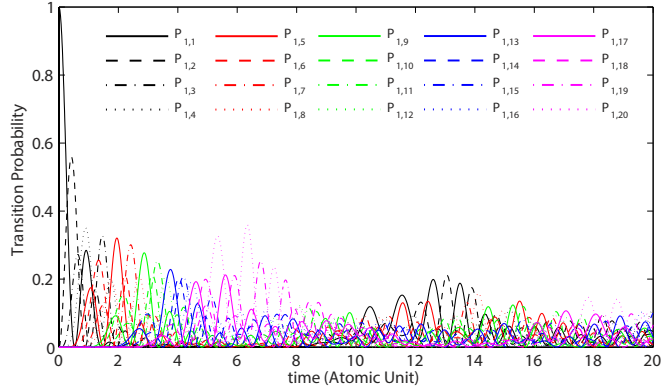}
    \end{center}
    \caption{(Color online) Transition probability from the first atomic state at time $t=0$ to 
    other states at time $t$.}
    \label{Fig-TransitionProbability}
    \end{figure}

Investigating charge transport in molecular junctions has been interesting for scientists 
~\cite{Nitzan,refcharge1,refcharge2,refcharge3,refcharge4}. Electrical transport properties of 
a system have a closed relation to the time evolution of the charge density in it. Consider a 
MMM system composed of a trans-polyacetylene molecule which contains 20 atoms as the central 
molecule and two $1D$ ideal semi infinite electrodes (figure~\ref{Fig-molecule2}), beside the 
assumption that the coupling energy between the polyacetylene molecule and each electrode is 
$v=0.5$ and an electric potential difference is applied between leads $\Delta V=2 volt$; the 
self-energy of each electrode can be calculated by equation~\eqref{EqSigmat} in the time 
representation. Hamiltonian of the trans-polyacetylene was written in the tight binding 
approximation and second quantization representation~\cite{Chien,Grant} as below:
      \begin{align}\label{Eq-PolyHam}
      H&=\sum_x \epsilon_0 c_x^\dagger c_x + \sum_{<x,x'>}^{'} t_{x,x'} c_x^\dagger      
      c_{x'}\nonumber\\
      t_{x,x'}&=t_0-2\alpha(-1)^{min(x,x')}\,u_0
      \end{align}        
In which the summations run over the lattice sites and the on-site energy of electrons 
($\epsilon_0$) is set properly. The hopping terms $t_{x,x'}$ relate to nearest neighbour sites. 
There are three empirical parameters which clarify the magnitude of the hopping integrals. The 
carbon-carbon atoms hopping integral related to $\pi$ orbitals equals to $t_0=2.5\,eV$. The 
electron-phonon coupling constant is $\alpha=4.1\,eV\AA^{-1}$ and $u_0=0.04\,\AA$ represents 
the constant displacement due to Peierls distortion because of 
dimerization~\cite{Peierls,Grant}. Applying this Hamiltonian and using the $FDM$, the 
equation~\eqref{cftgr} can be solved numerically.
Suppose $|\psi_i ,t=0>$ represents a state that an electron exists on $i$th atom (in the local 
atomic state $i$) at the time $t=0$. The TEO traces this state in the next time $t$ as follow:
     \begin{equation} \label{psiEvoltion}
      |\psi_i ,t>=  U(t)\,|\psi_i ,t=0>
     \end{equation}
then the probability density of transition from $i$th atom at time $t=0$ to $j$th atom at time 
$t$ is straightforward as:  
     \begin{equation} \label{probabilityEvoltion}
      P_{i,j}=  {|<\psi_j ,t|\psi_i ,t=0>|}^2
     \end{equation}
Evolution of this probability density in time may interpret as the movement of an electron 
wave packet between different atomic states in the system. For instance figure
\ref{Fig-TransitionProbability} shows the evolution of transition probability density from 
the first atomic state, $i=1$ at time $t=0$ to all other states at any time (equation
 \eqref{probabilityEvoltion}).

\begin{figure}[h]
    \begin{center}
    \includegraphics[scale=0.6]{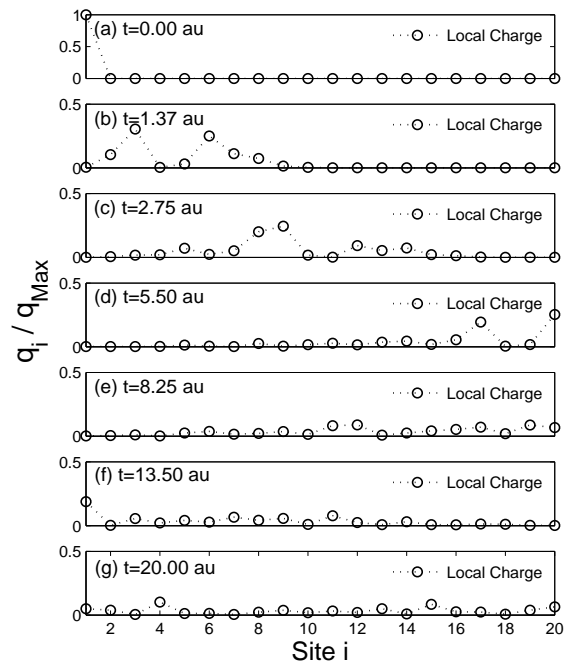}
    \end{center}
    \caption{The evolution of a charge density from left to right snapshots in the 
    atomic state basis.}
    \label{Fig-snapshots}
    \end{figure}

\begin{figure}[ht]
    \begin{center}
    \includegraphics[scale=0.6]{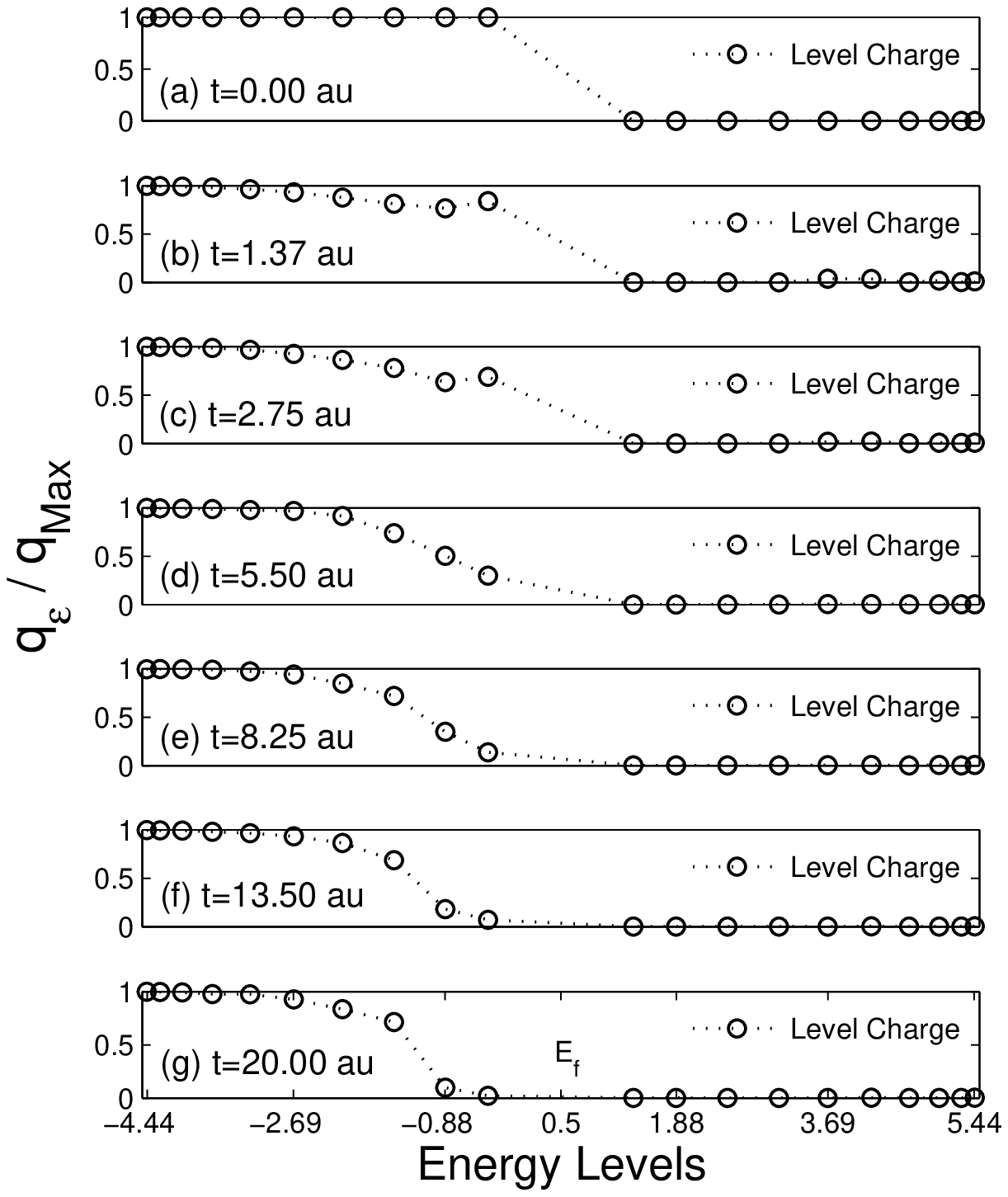}
    \end{center}
    \caption{The evolution of a charge density from left to right snapshots in the 
    energy basis.} 
    \label{Fig-Levelsnapshots}
    \end{figure}

To study charge transport in the system one needs to define the charge density in it. Assume     
$q_i(t)$ denotes the charge density of the atomic state $i$ at time $t$, which is related to 
initial charges on all states, $q_j(0)$ as:      
      \begin{equation} \label{qTime}
      q_i(t)=  \sum_j q_j(0) {|<\psi_i ,t|\psi_j ,t=0>|}^2
      \end{equation}
while its discrete counterpart can be rewritten in the matrix representation as:     
      \begin{equation} \label{VecqTime}
      \bold{q}^{\psi}(t)= \bold{P}^{\psi}(t)\,\bold{q}^{\psi}(0)
      \end{equation}      
In which $\bold{P}^{\psi}$ represents the matrix of the transition probabilities between 
atomic states and $\bold{q}^{\psi}$ is a vector containing the charge of each one. Applying the 
TEO on the charge density one may find the propagation of it in the system 
during a specific time. For instance assume, at the initial time ($t=0$) one electron is 
arrived in first atom of the molecule from left lead ($q_i(0)=\delta_{i,1}$). The snapshots of 
the $\bold{q}^{\psi}(t)$ at different times are shown in figure~\ref{Fig-snapshots}. At the 
time $t=5.5(Atomic\,Unit)$ this electron wave packet collides to the right lead and diffracts 
[see the figure~\ref{Fig-snapshots}(d)] . Some parts of it reflected back and others 
transferred to the right lead and this process persists in time. The time evolution of the 
Hamiltonian eigenstates is useful. Assume $|\epsilon>$ denotes the eigenstate with eigenvalue 
$\epsilon$. This means that initially the equation~\eqref{EigenState} holds for this state so $ 
|\epsilon,t=0> $ is calculated by: 
     \begin{equation} \label{EigenState}
      H(0)\,|\epsilon,t=0>= \epsilon\,|\epsilon,t=0>
      \end{equation}
then its time evolution can be calculated by:      
     \begin{equation} \label{EigenStateEvoltion}
      |\epsilon,t>=U(t)\,|\epsilon,t=0>
     \end{equation}
Suppose $q_{\epsilon}(t)$ denotes the charge density in energy level $\epsilon$ at time $t$. 
Rewriting equations~\eqref{qTime} and~\eqref{VecqTime} in these new basis, one may found: 
      \begin{equation} \label{LevelqTime}
      q_{\epsilon}(t)=\sum_{\epsilon'}q_{\epsilon'}(0){|<\epsilon,t|\psi_{\epsilon'},t=0>|}^2
      \end{equation}
and its discrete representation as:
      \begin{equation} \label{LevelVecqTime}
      \bold{q}^{\epsilon}(t)=\bold{P}^{\epsilon}(t)\,\bold{q}^{\epsilon}(0)
      \end{equation}      
In which $\bold{P}^{\epsilon}$ is a matrix which represents transition probabilities between 
energy eigenstates and $\bold{q}^{\epsilon}$ denotes a vector contains the charge of these 
energy levels. Consider half filling situation with sorted states as initial state at the time 
$t=0$ and define the Fermi energy at the middle of the energy difference between highest full 
state and lowest empty one, $E_F=0.5\,eV$. The snapshots of the time evolution of this initial 
state were shown in Figure~\ref{Fig-Levelsnapshots} in several times. During this process, 
states whose their energy are near the Fermi energy, lose their charge while others are robust. 
These lost charges are transferred to the leads and make a charge current. 

To calculate the current, definition of the total charge in the center part of the system at 
any time $t$ is mandatory. This can be properly defined by the summation of total charge in 
all atoms of central molecule at any time $t$, as: 
      \begin{equation} \label{qTotalTime}
      q_{T}(t)=  \sum_i q_i(t)
      \end{equation}

\begin{figure}[h]
    \begin{center}
    \includegraphics[scale=0.4]{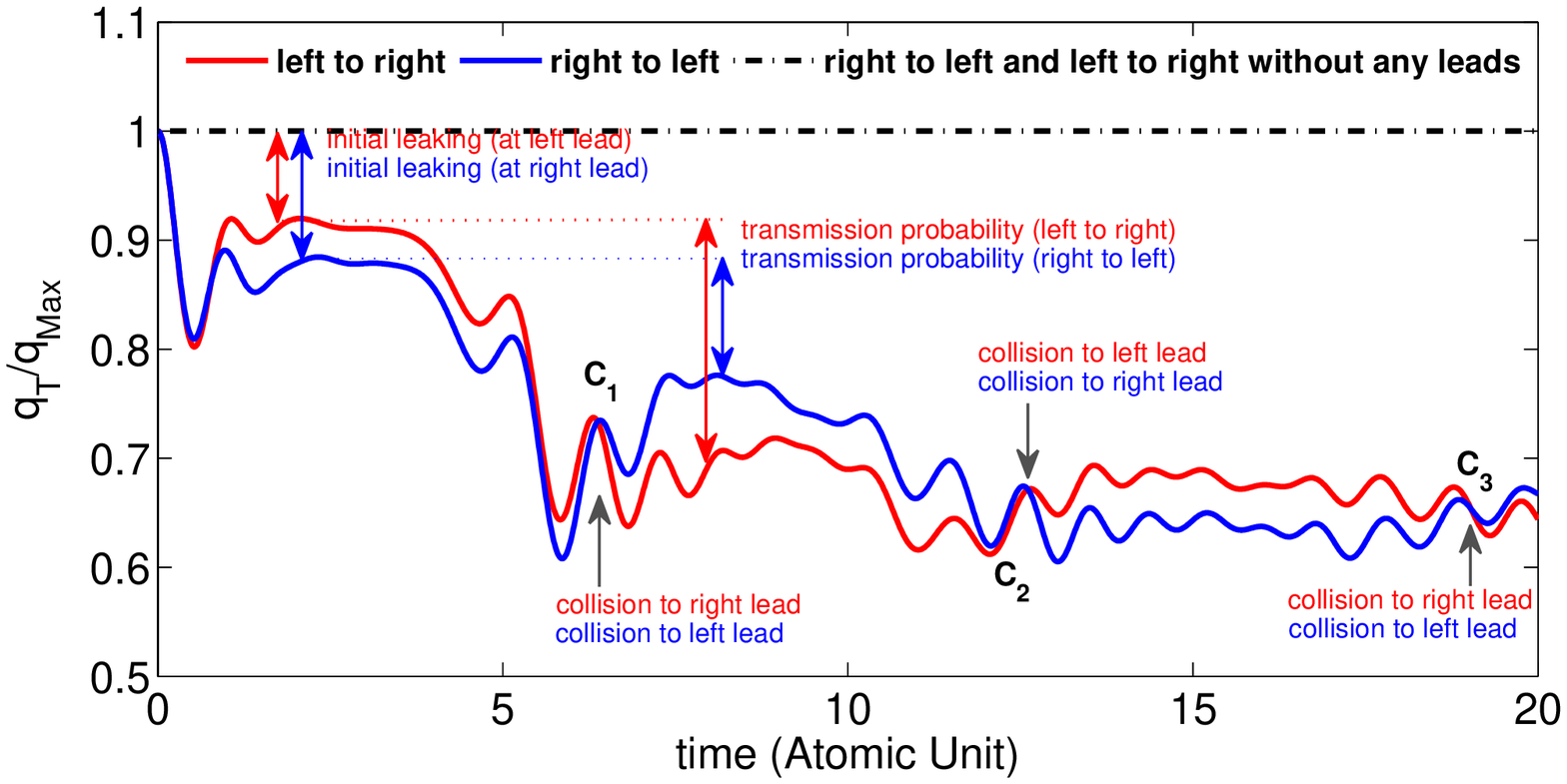}
    \end{center}
    \caption{(Color online) The total charge in molecule.}
    \label{Fig-qTotal}
    \end{figure}

\begin{figure}[ht]
    \begin{center}
    \includegraphics[scale=0.4]{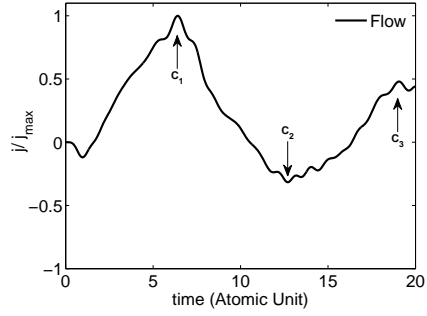}
    \end{center}
    \caption{The charge flow which is calculated by the energy eigenstates.} \label{Fig-Flow}
    \end{figure}

Evolution of this total charge in time illustrates the meaning of the charge current in the 
molecule. Three different cases were shown in figure~\ref{Fig-qTotal}. If the molecule become 
separated from the leads, its total charge must be constant which is compatible with the black 
dashed line diagram in figure~\ref{Fig-qTotal}. The red (blue) line diagram refers to the 
situation in which the electron is arrived in first (last) atom from left (right) lead at time 
$t=0$. Generally, amount of the charge reduces during the time evolution. At first this charge 
reduces rapidly and leaks to the leads but then partially increases due to contact effect. The 
potential difference between the leads causes the total charge magnitude for evolution from the 
left to the right of the molecule be grater than its reverse direction. When the maximum of the 
charge density collides to the next lead, these two lines are tangent to each others during the 
large reduction due to the charge leaking to the leads [compare the violet dot line in figure 
\ref{Fig-TransitionProbability} with figure~\ref{Fig-qTotal} at time $t=5.5(Atomic\,Unit)$] and 
therefore intersect where direction of the charge flow changes (points $c_1$, $c_2$ and $c_3$). 
The difference between the charge magnitude before and after charge leaking, determines net 
amount of charge transferred to the leads which makes a charge current.
Comparison between this current interpretation and formal one which relates to the current 
density operator is beneficial \{$j(t)=\frac{\hbar}{m}\Im[\psi^{\dagger}
(t)\nabla\psi(t)]$\}. The charge flow is plotted in figure \ref{Fig-Flow}. The extremums in 
this diagram are equal to collisions of the charge density with leads which is satisfiable.
Applying this method for the first and the last atoms of the molecule, we calculated the 
current. All the charge evolutions in these two atoms emerge from three distinct sources: 1. 
The initial charges which have remind there yet $q^{L(R)}(t)$, 2. The charges which depend on 
other sites $q^{L(R)\rightarrow R(L)}(t) $, 3. The interaction with the adjacent lead 
$q^{L(R)\rightarrow L_{l(r)}}(t)$. So we can find the total transferred charge from the first 
atom as:
      \begin{align}\label{qTotal11}
       q^{L}(t)&=P_{1,1}(t)\,q^{L}(0)\nonumber\\
       q^{L\rightarrow R}(t)&=\sum _{j=2}^{n}\,P_{1,j}(t)\,q^{L}(0)\nonumber\\
       q^{L\rightarrow L_l}(t)&=q^{L}(0)-q^{L}(t)-q^{L\rightarrow R}(t)
      \end{align} 
and so for the last atom we have:
      \begin{align}\label{qTotal11}
       q^{R}(t)&=P_{n,n}(t)\,q^{R}(0)\nonumber\\
       q^{R\rightarrow L}(t)&=\sum _{j=1}^{n-1}\,P_{n,j}(t)\,q^{R}(0)\nonumber\\
       q^{R\rightarrow L_r}(t)&=q^{R}(0)-q^{R}(t)-q^{R\rightarrow L}(t)
      \end{align} 
then the transferred charge from the molecule at any time $t$ is calculated as:
      \begin{equation} \label{Transcharge}
    q(t)=\frac{1}{2}\lbrace[q^{L\rightarrow R}(t)+q^{R\rightarrow L_r}(t)]-[q^{R\rightarrow L}
      (t)+q^{L\rightarrow L_l}(t)]\rbrace 
      \end{equation} 
As our computations are in the \textit{independent electron approximation}, total transferred 
charge at every time $t$, is sum of all transferred charges in any infinitesimal time period 
$\delta\tau$. So we have: 
      \begin{equation} \label{Totalcharge}
      q(t)^T=q(t)+q(t-\delta\tau)+q(t-2\delta\tau)+\cdots\simeq\int_0^t\,q(t-\tau)\,d\tau
      \end{equation}
    \begin{figure}[h]
    \begin{center}
    \includegraphics[scale=0.4]{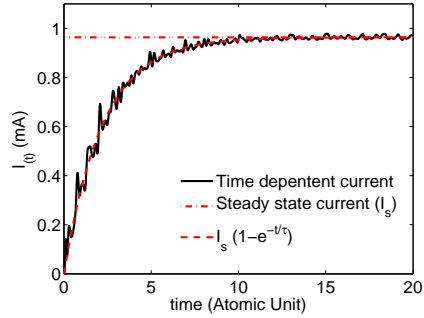}
    \end{center}
    \caption{(Color online) The charge current as a function of time ($t$) in the system at a 
    constant voltage}\label{Fig-It}
    \end{figure}
    \begin{figure}
    \begin{center}
    \includegraphics[scale=0.43]{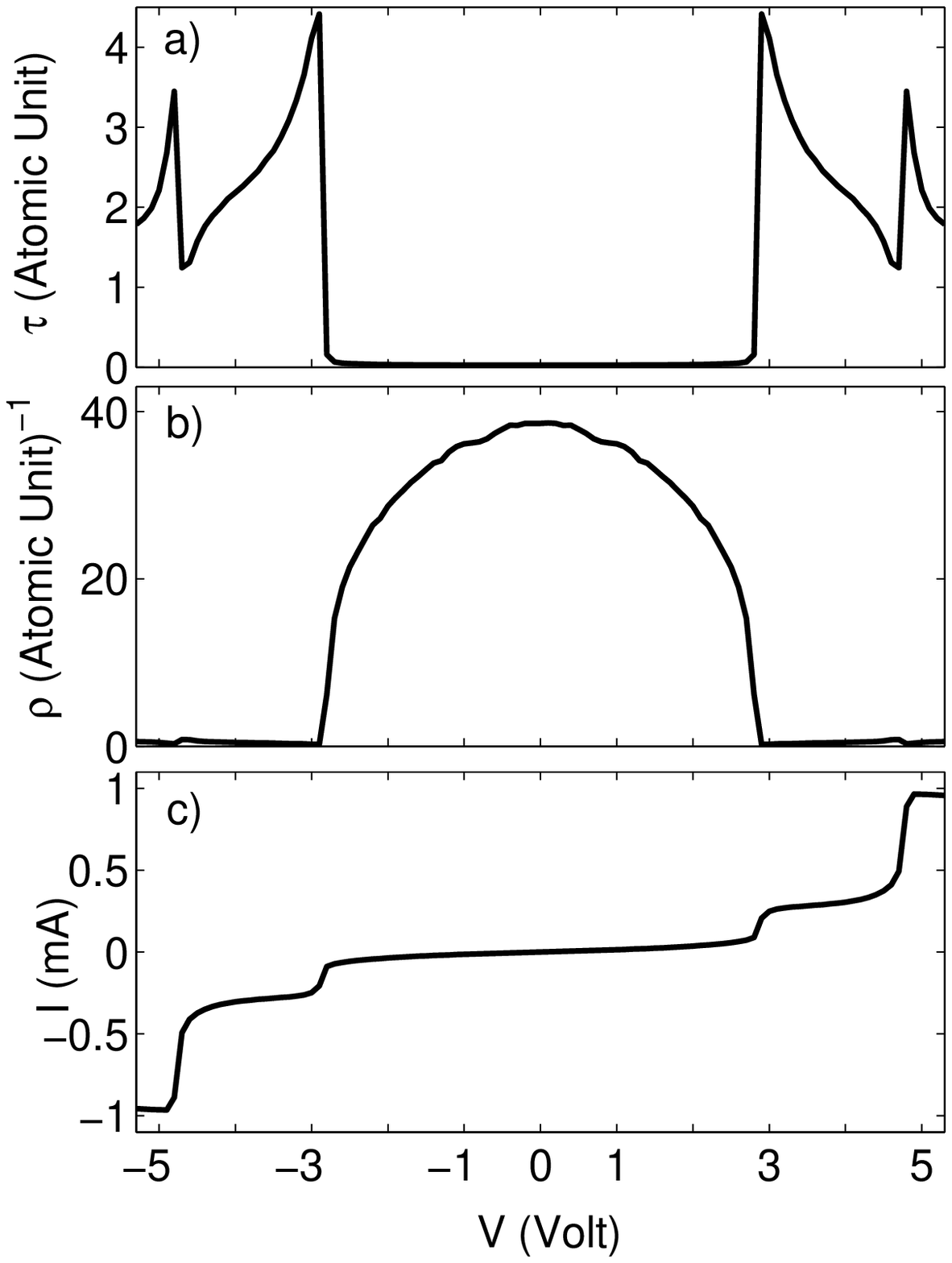}
    \end{center}
    \caption{The relaxation time-voltage (a) the resistivity-voltage (b) and the current-
    voltage (c) diagrams.}
    \label{Fig-IVSet}
    \end{figure}

Therefore its time derivation will give the current [$I(t)=\frac{d}{dt}q(t)^T$]. After a proper 
time, for every constant voltage an steady current $I_s$, will pass trough the system. 
Figure~\ref{Fig-It} shows the current diagram at constant voltage $v=5\,V$ 
where the steady current is $I_s=0.96\,mA$. This tranquility time may interpret as a relaxation 
time, $\tau$, which can be calculated by fitting the diagram of figure~\ref{Fig-It} to a proper 
function like $f(t)=1-e^{-\frac{t}{\tau}}$. The inverse of this relaxation time ($\tau^{-1}$) 
may be regarded as resistivity ($\rho$) of this system. 

Following this procedure for other voltages, we can find the voltage dependent diagrams of 
these three parameters. Indeed the relaxation time-voltage, the resistivity-voltage and the 
current-voltage diagrams are found as depicted in figures~\ref{Fig-IVSet} (a), (b) and (c), 
respectively. Existence of steps in the current-voltage diagram, is a proper evidence for 
quantum confinement effect. 

\section{Summery}
\label{sec4}
In summery, we propose a new numerical method to study time evolution in physical systems by 
using $FDM$. To solve the correspondent $Volterra\,integro-differential\,equation$, first we 
introduced a first order derivative and an integrator operators and discretized them. Using 
this method we studied the time evolution of a $1D$ chain Hamiltonian in different situations 
and compared our results with Dyson series and Runge Kutta. Our method not only is compatible 
with analytical results but also is more accurate than other numerical methods. Furthermore we 
study the charge transport in a trans-polyacetylene chain as a central molecule of a $MMM$ 
system by considering time evolution of its charge density and then calculated its current-
voltage diagram.

The most significant application emerges from this method that has not instantly mentioned is 
that it can properly be applied for time dependent Hamiltonians regardless of the source of 
this time dependency. So it not only can be used for time dependent Hamiltonian but also may be 
used for time dependent self-energies related to the electrodes in $MMM$ system.



\end{document}